\documentclass[reprint,prr,twocolumn,twoside,aps]{revtex4-1}

\usepackage{amsmath,amssymb,amsfonts}
\usepackage{graphicx}
\usepackage{siunitx}
\usepackage{natbib}
\usepackage{braket}
\usepackage{multirow}
\usepackage[normalem]{ulem}
\usepackage{xspace}
\usepackage{url}
\usepackage[breaklinks]{hyperref}
\hypersetup{
    unicode=true,
    colorlinks=true,
    linkcolor=blue,
    citecolor=blue,
    urlcolor=blue
  }
	\usepackage{breakurl}
	\usepackage[dvipsnames]{xcolor}

\newcommand{\Yb}{$^{173}$Yb\xspace}

\usepackage{xr}

\begin{document}

\title{Observation of magnetic Feshbach resonances between Cs and \Yb}

\author{Tobias Franzen$^1$}
\thanks{These three authors contributed equally.}
\author{Alexander Guttridge$^1$}
\thanks{These three authors contributed equally.}
\author{Kali E. Wilson\textsuperscript{1,2}}

\thanks{These three authors contributed equally.}
\author{Jack Segal$^1$}
\author{Matthew D. Frye$^3$}
\author{Jeremy M. Hutson$^3$}
\email{j.m.hutson@durham.ac.uk}
\author{Simon L. Cornish$^1$}
\email{s.l.cornish@durham.ac.uk}

\affiliation{$^1$Joint Quantum Centre (JQC) Durham-Newcastle, Department of Physics, Durham University, South Road, Durham, DH1 3LE, United Kingdom.}
\affiliation{$^2$Department of Physics, SUPA, University of Strathclyde, Glasgow G4 0NG, United Kingdom}
\affiliation{$^3$Joint Quantum Centre (JQC) Durham-Newcastle, Department of Chemistry, Durham University, South Road, Durham, DH1 3LE, United Kingdom.}

\begin{abstract}
We report the first observation of magnetic Feshbach resonances between \Yb and $^{133}$Cs. In a mixture of Cs atoms prepared in the $(f=3, m_f=3)$ state and unpolarized fermionic \Yb we observe resonant atom loss due to two sets of magnetic Feshbach resonances around 622~G and 702~G. Resonances for individual Yb nuclear spin components $m_{i,\mathrm{Yb}}$ are split by its interaction with the Cs electronic spin, which also provides the main coupling mechanism for the observed resonances. The observed splittings and relative resonance strengths are in good agreement with theoretical predictions from coupled-channel calculations.
\end{abstract}
\date{\today}

\maketitle
\section{Introduction}

Magnetic Feshbach resonances are an essential tool for controlling the effective contact interactions between atoms and molecules. Their application in the field of ultracold gases has led to many exciting discoveries \cite{Chin2010}. The exquisite control of atomic interactions offered by Feshbach resonances has been instrumental in studies of three-body physics \cite{Kraemer2006,Naidon2017}, the BEC-BCS crossover in Fermi gases \cite{Regal2004,Zwierlein2004} and the creation of quantum droplets \cite{Petrov2015,Kadau2016,Cabrera2018, DErrico2019}. They also allow the formation of ultracold molecules by magnetoassociation \cite{Koehler2006,Hutson2006}, where the magnetic field is adiabatically ramped across a Feshbach resonance to create a weakly bound molecule. The Feshbach molecules formed in this process can then be transferred to their absolute ground states by stimulated Raman adiabatic passage \cite{Bergmann1998}. This protocol for molecule association has been widely employed for molecules formed from two alkali atoms \cite{Ni2008,Lang2008,Danzl2008,Takekoshi2014,Molony2014,Park2015,Guo2016,Rvachov2017,Seeselberg2018a,Yang261,Voges2020b,Zhang2020}.

Recently, there has been much interest in atomic mixtures composed of alkali-metal and closed-shell atoms \cite{Tassy2010,Hara2014,Pasquiou2013,Khramov2014,Vaidya2015,Guttridge2017,Flores2017,Witkowski2017, Ye2020}. The contrasting electronic structures of the constituent atoms offer opportunities to perform species-specific manipulations through application of optical or magnetic fields. The ability to perform low-cross-talk, species-specific manipulations on these mixtures will allow studies of impurity physics and topological superfluids in mixed dimensions \cite{Wu2016, Loft2017, Schafer2018} and enhance studies of collective dynamics \cite{Modugno2002a, Ferrier2014, Roy2017, DeSalvo2019,Wilson2021a}, and vortex interactions in quantum mixtures \cite{Yao2016, Kuopanportti2018}. Such mixtures also offer an avenue for the creation of $^{2}\Sigma$ molecules which, due to an unpaired electron spin, exhibit a magnetic dipole moment. The presence of a magnetic dipole moment, in addition to an electric dipole moment, provides an extra degree of tunability to the system, and will allow simulation of a greater variety of spin Hamiltonians \cite{Micheli2006} and exploration of new quantum phases \cite{Perez-Rios2010}. In addition, these molecules offer further prospects for controllable quantum chemistry \cite{Abrahamsson2007,Quemener2016} and quantum computation \cite{Herrera2014,Karra2016}.

In order for alkali--closed-shell mixtures to realise their full potential, a magnetic Feshbach resonance must be observed between the constituent atoms of the mixture. The observation of Feshbach resonances and the formation of $^{2}\Sigma$ molecules using these mixtures is challenging, however, because of the narrow expected widths of the Feshbach resonances. The absence of electronic spin in the $^{1}$S ground state of the closed-shell atom removes the strong couplings which produce the wide resonances present in bi-alkali systems. Narrow resonances are predicted to occur in alkali--closed-shell mixtures, either through modification of the hyperfine coupling of the alkali atom caused by the closed-shell atom at short range \cite{Zuchowski2010,Brue2013}, or by interaction between the electron spin of the alkali atom and the closed-shell atom \cite{Brue2012}. These narrow resonances have been experimentally observed, first in Rb + Sr \cite{Barbe2018} and later in Li + Yb \cite{Green2020}. However, magnetoassociation of $^{2}\Sigma$ molecules by sweeping the magnetic field across these resonances is yet to be achieved.

Yang \emph{et al.}\ \cite{Yang2019} recently carried out a theoretical investigation of Feshbach resonances in Cs + Yb. They made specific predictions of resonance positions and widths for all stable isotopes of Yb. The resonance positions are based on a ground-state electronic potential determined principally from the binding energies of near-threshold molecular states, measured by 2-photon photoassociation spectroscopy \cite{Guttridge2018a}. In Cs + Yb, the seven stable isotopes of Yb and the large mass of Cs allow significant tuning of the atom-pair reduced mass by using different Yb isotopes. Coupled with the different spin statistics of the Yb isotopes (five spin-zero bosons and two fermions), this leads to a wide variety of Feshbach resonances with different properties. Notably, the Bose-Fermi pairings of Cs + \Yb and Cs + $^{171}$Yb possess the densest Feshbach spectrum and the widest resonances. These resonances are mostly due to coupling mechanisms that arise from the non-zero nuclear spin of the fermionic Yb isotopes. Unfortunately, the preparation of mixtures of Cs and $^{171}$Yb is difficult due to the small \emph{intra-}species scattering length of $^{171}$Yb ($a_{\mathrm{Yb}}=-3 \, a_{0}$) \cite{Kitagawa2008}. This necessitates sympathetic cooling of $^{171}$Yb using another species, which is not compatible with our current protocol for preparation of Cs + Yb mixtures \cite{Wilson2021}.We therefore focus on the experimentally more accessible combination of Cs + \Yb.

\begin{figure*}[t] 
\begin{center}\includegraphics[width=1\textwidth]{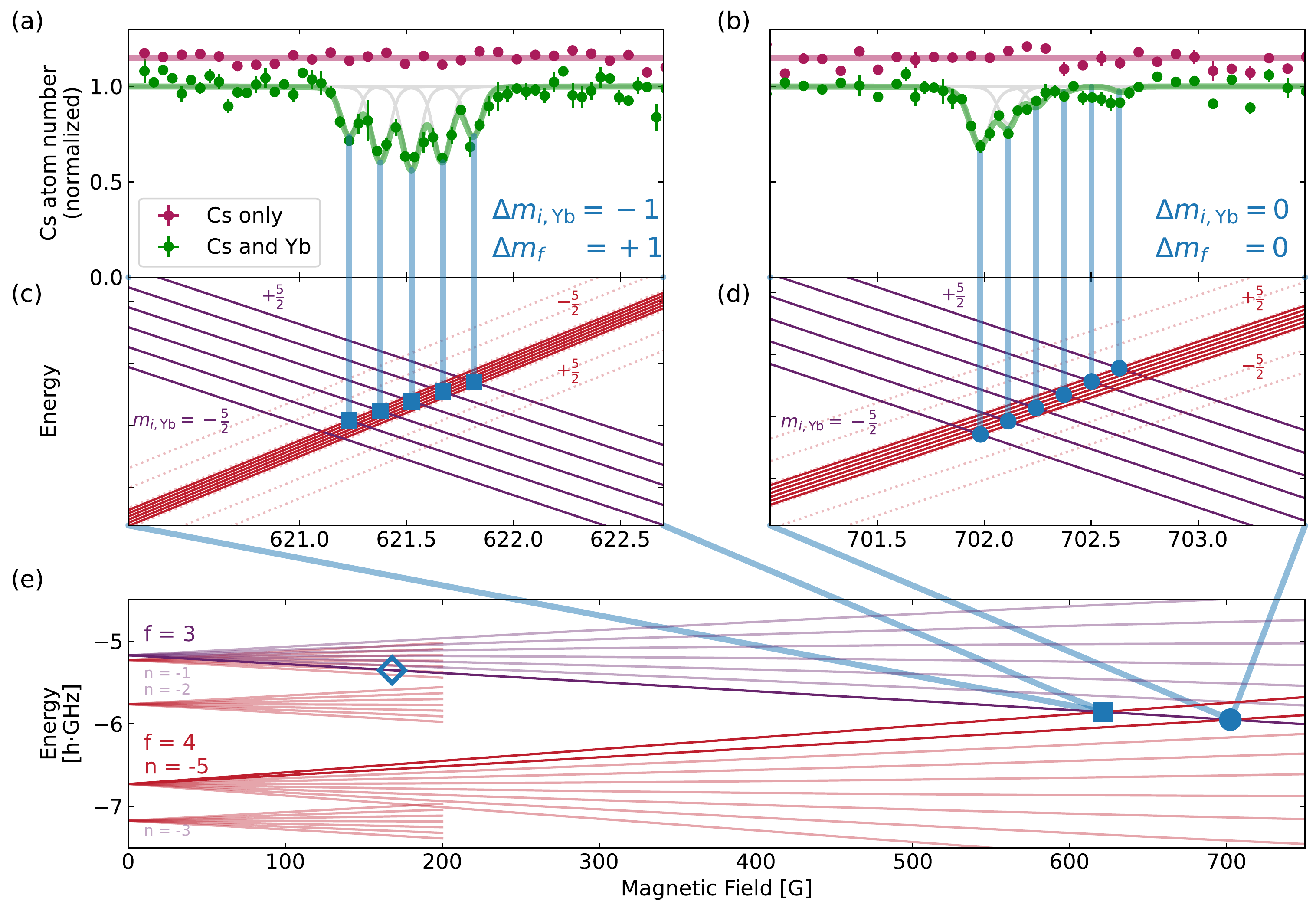}\end{center}
\caption{Origin and detection of Feshbach resonances for Cs+\Yb. \textbf{(a,b)} Relative number of Cs atoms remaining after holding for 1~s at a variable magnetic field. Green (red) points show measurements performed with (without) \Yb in the dipole trap. The normalised Cs number in the absence of \Yb is offset for clarity. Solid lines are fitted sums of Voigt profiles, with the individual components shown in gray.
\textbf{(c,d)} Level crossings giving rise to the detected resonances for the individual spin states labeled by $m_{i,\mathrm{Yb}}$. Symbols indicate crossings fulfilling $\Delta m_{i,\mathrm{Yb}} = -\Delta m_f$. Faint dotted lines show the bound states without the shifts induced by mechanism~II, in which case all resonances would occur at the same field. The vertical scale is $h\cdot4\,\mathrm{MHz}$.
\textbf{(e)} Relevant  thresholds (purple lines) and  bound-state energies (red lines)  over the experimentally accessible field range. Vibrational levels that do not give rise to an observed resonance are truncated to low fields for clarity. Structure due to different values of $m_{i,\mathrm{Yb}}$ is not visible on the scale of \textbf{(e)}.
\label{fig:FR}}
\end{figure*}

In this paper, we report the first observation of magnetic Feshbach resonances between Cs and Yb atoms. We prepare a mixture of Cs and \Yb with high phase-space density and then perform atom loss spectroscopy to observe a magnetic Feshbach resonance. The paper is organised as follows:
In Sec. \ref{sec:FR} we briefly introduce the underlying theory. In Sec. \ref{sec:exp} we describe our experimental procedure for preparing an ultracold mixture of Cs and \Yb and performing Feshbach spectroscopy.
In Sec. \ref{sec:results} we use these methods to detect two sets of Feshbach resonances and confirm theoretical predictions of the distance-dependent hyperfine coupling by measuring their splitting.

\section{Feshbach resonances between Cs and \Yb}\label{sec:FR}

The general structure of the near-threshold energy levels predicted for Cs$^{173}$Yb \cite{Yang2019} is shown in Fig.~\ref{fig:FR}. Each threshold of Cs, labeled by total spin $f$ and projection $m_f$ onto the axis of the field, supports near-threshold bound states labelled by vibrational quantum number $n$, where $n=-1$ is the least-bound state for partial wave $L=0$. The feature most relevant for the present work is the set of bound states with $n=-5$ supported by the Cs hyperfine thresholds with $f=4$. At zero field these states lie about 1.56 GHz below the thresholds for $f=3$. For $^{173}$Yb, both bound states and thresholds are further split into components labelled by $m_{i,\textrm{Yb}}$, the projection of the \Yb nuclear spin $i_\textrm{Yb}=5/2$. 

There are three different mechanisms that provide coupling between the bound states and the thresholds they cross. First, the Cs nuclear hyperfine coupling constant $\zeta_\textrm{Cs}$ is modified by a quantity $\Delta\zeta_\textrm{Cs}(R)$ when an Yb atom approaches to a distance $R$ \cite{Zuchowski2010}; this is referred to as mechanism I, and provides couplings between bound states and thresholds with the selection rule $\Delta m_f=0$. Secondly, the Yb atom experiences a hyperfine coupling $\Delta\zeta_\textrm{Yb}(R)$ due to the transfer of part of the electronic spin of Cs onto the Yb atom \cite{Brue2012}; this is referred to as mechanism II, and provides couplings between bound states and thresholds with the selection rule $\Delta m_f=0,\pm1$. Mechanism III, due to tensor hyperfine couplings \cite{Barbe2018}, is not important for the present work.

\section{Experimental Setup}\label{sec:exp}

In our setup, we detect magnetic Feshbach resonances between Cs and \Yb through enhanced loss of Cs atoms. In order to observe Feshbach resonances using this technique we must first prepare an ultracold mixture of Cs and $^{173}$Yb. Our experimental methods for preparing ultracold mixtures of Cs and Yb are detailed in earlier works \cite{Kemp2016,Hopkins2016,Guttridge2016,Guttridge2017,Guttridge2018,Guttridge2018a,Wilson2021}. 
As in our previous investigations of Cs+Yb photoassociation \cite{Guttridge2018,Guttridge2018a}, the experiments reported here are performed in thermal mixtures. However, the addition of a bichromatic optical dipole trap (BODT), as reported in Ref. \cite{Wilson2021}, results in improved control over the overlap of the two species. 

The BODT is formed of three beams with two different wavelengths. A single $532\,\mathrm{nm}$ beam with beam waist $50(3)\, \mu \mathrm{m}$ co-propagates with a $1070\,\mathrm{nm}$ beam with a $33(3)\, \mu \mathrm{m}$ beam waist. Another $1070\,\mathrm{nm}$ beam of waist $72(4)\, \mu \mathrm{m}$ crosses the $532\,\mathrm{nm}$ and $1070\,\mathrm{nm}$ beams at an angle of $40^{\circ}$ to create a crossed beam trap. 

To prepare the Cs + \Yb mixture in the BODT we first load an \Yb MOT and then transfer it into the BODT.
The \Yb sample is then evaporatively cooled by reducing the power of the BODT over 3~s. During this Yb evaporation stage we load the Cs MOT. The BODT powers during the evaporation are such that the overlap region of the three BODT beams produce a potential which is repulsive for Cs (yet attractive for Yb); we do not observe any deleterious effects from loading the Cs MOT alongside the Yb sample, probably because of the small overlap density of the two clouds. At the end of the Yb evaporation, the BODT powers are modified to produce a trap suitable for loading Cs, while maintaining a similar trap depth for $^{173}$Yb. Simultaneously, the Cs MOT is overlapped with the BODT and cooled using optical molasses, followed by degenerate Raman sideband cooling (DRSC). DRSC cools the Cs atoms to $T \approx 1 \, \mu$K and optically pumps them into the hyperfine state $(f=3,m_f=3)$. Minimal loading of Cs atoms into the BODT is observed at this stage, so in order to enhance the efficiency of the transfer into the BODT we first transfer Cs into a large-volume dipole trap. During this stage we apply a magnetic field gradient of 31.3\,G/cm to levitate the atoms \cite{Li2015} and a magnetic bias field of 90~G to increase the Cs elastic collision rate. Following a 500~ms hold, the bias field is ramped down to the Efimov minimum in the three-body recombination rate at 22~G \cite{Kraemer2006} and the magnetic field gradient is ramped to zero. The trapping light for the large-volume trap is then switched off, leaving a sample of Cs and Yb confined in the BODT. The powers of the BODT are then ramped to their final value in order to prepare the atomic mixture at the appropriate temperatures and densities for the Feshbach resonance measurements. This results in a trap with geometric mean trapping frequencies $\bar{\nu}_{\rm Cs}  = 254 \, \mathrm{Hz}$ for Cs and $\bar{\nu}_{\rm Yb}  = 107 \, \mathrm{Hz}$ for Yb. In our previous work with \textsuperscript{174}Yb \cite{Wilson2021a}, sympathetic cooling allowed the preparation of dual-degenerate samples. Interspecies thermalization is observed for Cs+\Yb despite the low interspecies scattering length of  of $a_\mathrm{CsYb} = 1(1)\, a_0$ \cite{Guttridge2018a}. However, it is far too slow for efficient sympathetic cooling, so we are currently restricted to working with thermal samples.

Once the Cs + \Yb mixture is prepared, the magnetic bias field is ramped to the desired value in 12~ms. The atomic mixture is held for 1~s at the desired magnetic field before quickly ramping down the magnetic field to near zero and performing dual-species absorption imaging to measure the remaining atom number of each species. Further details on the setup and calibration of the magnetic field are given in the Appendix.

\section{Results}\label{sec:results}

A typical experiment to measure an interspecies Feshbach resonance starts with an unpolarized sample of $1 \times 10^{5}$ \Yb atoms equally distributed over the six nuclear spin states with $m_{i,\textrm{Yb}}$ from $-5/2$ to $+5/2$ at a temperature $T= 0.8 \, \mu$K and $3 \times 10^{5}$ Cs atoms in the state $(f=3, m_f=3)$ at a temperature $T=4 \, \mu$K. Only very slow thermalization of the sample, on the timescale of seconds, is observed away from resonance, which is to be expected given the interspecies scattering length of $a_\mathrm{CsYb} = 1(1)\, a_0$ \cite{Guttridge2018a}. 

Three-body loss is typically a useful process in searches for Feshbach resonances, because the three-body loss rate is strongly enhanced for the large scattering lengths which occur near the pole of a Feshbach resonance. However, Cs+Cs has both a rich Feshbach spectrum and a large background scattering length, so there are many magnetic fields where $a_{\rm Cs}$ is large \cite{Berninger2013}. At the magnetic fields explored in these measurements we observe a large amount of background Cs loss due to the high (homonuclear) three-body recombination rate. 
This fast loss of Cs due to Cs+Cs+Cs three-body collisions poses a challenge for the detection of an interspecies Feshbach resonance, as it has the potential to obscure any Cs loss due to Cs+\Yb+\Yb or Cs+Cs+\Yb collisions. The Cs three-body recombination rate is very large around 600~G, with a value of $K_{3,\mathrm{Cs}} \approx 1 \times 10^{-23} \, \mathrm{cm}^{6} \, \mathrm{s}^{-1}$ predicted in the zero-energy collision limit \footnote{Using Eq.~(3) from Ref.~\cite{DIncao2004}, $\eta=0.06$ and $a_{+}=1060 \, a_{0}$ \cite{Kraemer2006} and the scattering lengths predicted from the best Cs potential in Ref. \cite{Berninger2013}.}.

We observe that after a 1~s hold time there is less than 5~\% of the initial Cs atom number remaining in the absence of Yb. This fast reduction in the Cs density occurs as we increase the magnetic field away from the Cs Efimov minimum. Therefore, although we start with similar atom numbers of Cs and Yb, a larger number imbalance in favor of Yb quickly manifests. This necessitates using Cs as a probe for the detection of any Cs+Yb resonance because the fractional change in the Yb number would be very small, particularly in an unpolarized sample of \Yb where approximately only  one sixth of the population contributes to the signal at a given field. 

\autoref{fig:FR} shows the origins and the detection of two sets of Cs+\Yb Feshbach resonances. Near the Feshbach resonances there is an enhancement in the three-body recombination rate between Cs and Yb atoms which causes a decrease in the number of Cs atoms retained in the dipole trap, as shown in Fig.~\ref{fig:FR}a and \ref{fig:FR}b. 
The Feshbach spectrum of Cs+Cs in the $(f=3,m_f=3)$ state was well characterised in Ref.~\cite{Berninger2013} and the loss features in the Cs atom number measured here cannot be explained by any predicted or observed Cs+Cs Feshbach resonances. To demonstrate further that the resonances observed are indeed interspecies Feshbach resonances, we perform a similar experiment using Cs alone. We prepare the Cs gas at broadly the same temperature and density so as to maintain the same sensitivity to resonances. As expected, the resonant loss features disappear when there is no Yb present. 

The observed interspecies resonances arise due to bound states with $f=4$ and $n=-5$ crossing the threshold for $(f=3,m_f=3)$ around 622~G with $\Delta m_f = + 1$ and around 702~G with $\Delta m_f =0$, as illustrated in Fig.~\ref{fig:FR}e. 
We note a third set of resonances associated with the same vibrational state but with $\Delta m_f = -1$ is expected around $806\,\mathrm{G}$, but this unfortunately lies beyond our experimentally accessible field range and the resonances are predicted to be weaker by a factor of $\sim 50$.

Each set consists of multiple resonances, corresponding to different values of $m_{i,\mathrm{Yb}}$, with $\Delta m_{i,\mathrm{Yb}} = -\Delta m_f$. These resonances are split by the diagonal matrix elements associated with the mechanism~II coupling term $\Delta\zeta_\mathrm{Yb}(R) \hat i_\mathrm{Yb}\cdot \hat s$, leading to 5 distinct resonances for $\Delta m_f = +1$ as shown in Fig.~\ref{fig:FR}c and 6 distinct resonances for $\Delta m_f = 0$ as shown in Fig.~\ref{fig:FR}d.

In our experiment the observed width of the resonances is dominated by magnetic field noise rather than the intrinsic widths of the resonances. The data are thus fitted \footnote{The free fit parameters are individual center frequencies and amplitude scalings for each set and a common scaling factor for the predicted splittings.} by a sum of Voigt profiles with Gaussian widths of 0.04~G as independently determined from microwave spectroscopy on Cs atoms.
For the center positions of the sets, corresponding to a fictional component with $m_{i,\mathrm{Yb}} = 0$, we obtain $(621.45 \pm 0.01_\mathrm{stat}\pm 0.03_\mathrm{syst})\,\mathrm G$ for the set with $\Delta m_f =+1$ and $(702.21 \pm 0.01_\mathrm{stat}\pm 0.03_\mathrm{syst})\,\mathrm G$ for the set with $\Delta m_f =0$. These may be compared with the values of 619.46~G and 700.12~G predicted in Ref.\ \onlinecite{Yang2019}. The shifts of $\sim 2$~G between the predicted and experimental positions are probably be due to uncertainties in the measured binding energies and to the neglect of Cs hyperfine shifts in obtaining the ground-state interaction potential of Ref.\ \onlinecite{Guttridge2018a}. The separation between the two sets depends mostly on the precisely known Zeeman levels of Cs, but has a small contribution from $\Delta \zeta_\textrm{Cs}(R)$, shifting the $\Delta m_f = 1$ resonance down by about 0.04~G, comparable to our experimental uncertainty.

\begin{figure}[tb] 
\includegraphics[width=1\columnwidth]{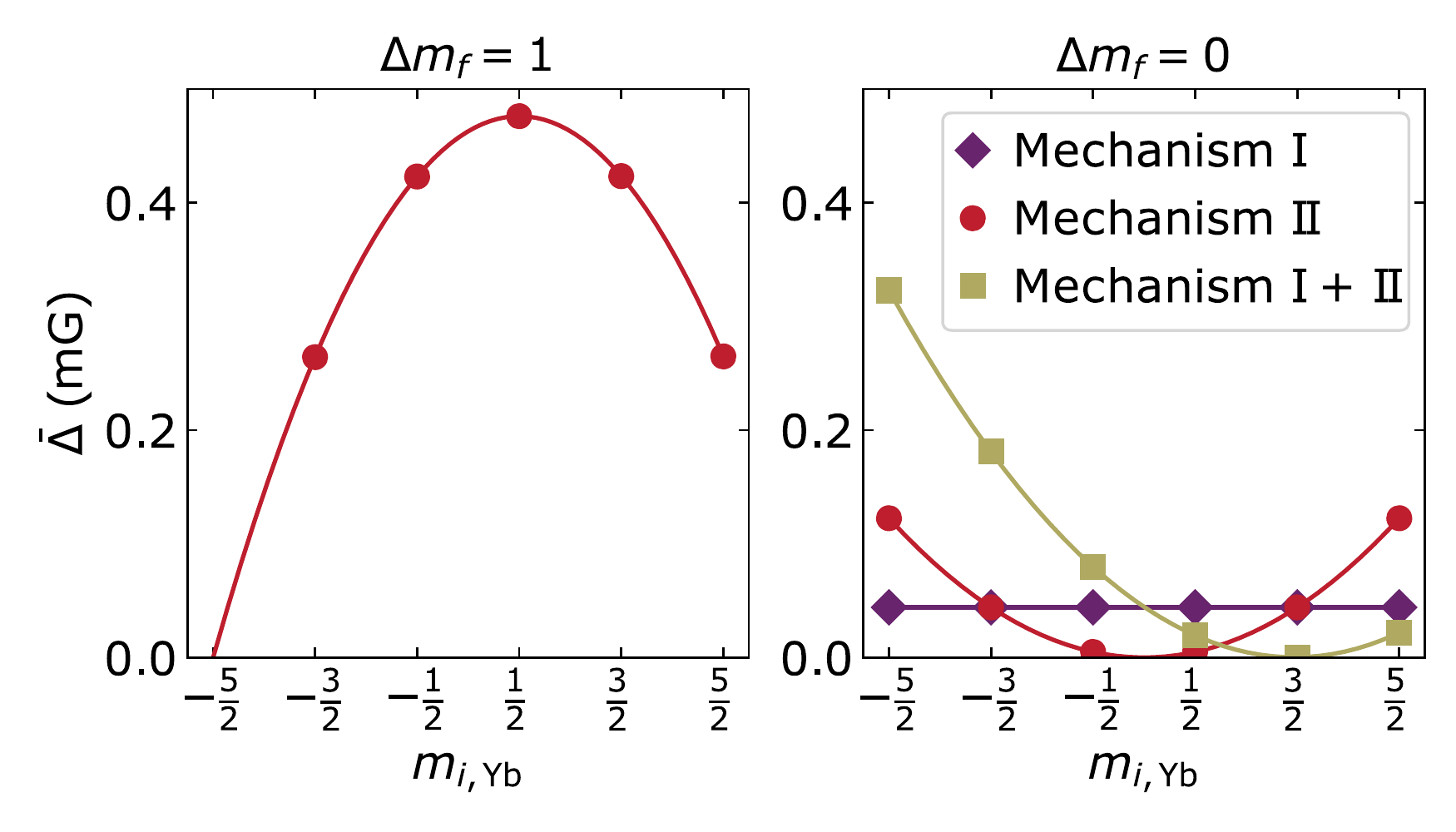}
\caption{Normalized resonance widths $\bar{\Delta}$ for the two observed sets of resonances with $\Delta m_f = 1$ and $\Delta m_f = 0$ \cite{Yang2019}. For $\Delta m_f = 1$ only mechanism~II contributes, while for $\Delta m_f = 0$ both mechanism~I and mechanism~II contribute.
	\label{fig:widths}}
\end{figure}    

The spacing between individual spin components is determined to be  $0.144(2)\,\mathrm G$ and  $0.131(2)\,\mathrm G$ for the 622\,G and 702\,G resonances respectively, where we constrain the fit to keep their ratio fixed at the theoretical value of $0.91$. These observed splittings are within 15~\% of the predicted values of 0.16~G and 0.14~G \cite{Yang2019}, implying that the calculated matrix elements of $\Delta\zeta_\mathrm{Yb}(R)$ are similarly accurate.

The resonance strengths within each set are predicted to show a parabolic dependence on $m_{i,\mathrm{Yb}}$ \cite{Yang2019}. For $\Delta m_f = +1$, the coupling is entirely due to mechanism~II, and the width peaks at $m_{i,\textrm{Yb}} = \frac 1 2$. For $\Delta m_f = 0$, both mechanism~I and mechanism~II couplings contribute.   The mechanism~I matrix element is constant across the set but the mechanism~II matrix element changes sign with $m_{i,\mathrm{Yb}}$; this is predicted to produce almost complete cancellation for the component with $m_{i,\mathrm{Yb}} = \frac 3 2$.
\autoref{fig:widths} shows the normalized widths of Yang et al. \cite{Yang2019} for these resonances;
their relative values are used to calculate the green lines in Fig.~\ref{fig:FR}a and b. It may be seen that the experimental profiles are in good agreement with the predictions, confirming the calculated ratio between the matrix elements for mechanism~I and mechanism~II \cite{Yang2019}.

A further interspecies resonance in our accessible field range is expected around $167\,\mathrm{G}$,  arising from bound states with $(f=3, m_f =2, n=-1)$ and indicated by an open diamond in Fig.~\ref{fig:FR}e. This resonance holds the promise of further insights into the distance dependence of $\zeta_\textrm{Yb}(R)$ due to the longer-range vibrational wavefunction of the bound state with $n=-1$, but it is unfortunately not observable under the current experimental conditions due to its smaller width and lower differential magnetic moment.

\section{Conclusion}\label{sec:conclusion}
We have reported the first observation of magnetic Feshbach resonances in a mixture of Cs and Yb. Each resonance is split into multiple components according to the nuclear spin state of \Yb. The resonance positions and the splittings and relative strengths of the components agree well with the predictions of Yang \emph{et al.}\ \cite{Yang2019}. This confirms the accuracy of the distance-dependent hyperfine couplings $\Delta\zeta_\mathrm{Cs}(R)$ and $\Delta\zeta_\mathrm{Yb}(R)$ obtained in Ref.\ \onlinecite{Yang2019} and consequently the predicted resonance widths for all isotope combinations. 

The ability to control the atomic interactions in this mixture using a Feshbach resonance offers many new avenues of research. These include the study of tunable Bose-Fermi mixtures in mixed dimensions \cite{Lamporesi2010,Wu2016, Caracanhas2017,Loft2017} using species-selective trapping potentials. The resonance may allow access to the strong-coupling regime \cite{Lewenstein2004a} in a way that is not feasible with the small-to-moderate interspecies background scattering lengths of these systems. It may also allow the implementation of a three-body hardcore constraint in optical lattices through induced three-body loss \cite{Mark2012,Diehl2010}.

The observation of magnetic Feshbach resonances in Cs+Yb is a vital step towards the production of CsYb molecules in the $^{2}\Sigma$ state through magnetoassociation. The small interspecies scattering length of the Cs+\Yb is advantageous here as it ensures miscibility \cite{Papp2008, Wang2015} and will allow the preparation of a mixed Mott insulator phase \cite{Sugawa2011} as an ideal starting point for magnetoassociation. The molecules could then be transferred to the rovibrational ground state using stimulated Raman adiabatic passage
and offer new opportunities for quantum simulation \cite{Micheli2006}, novel quantum phases \cite{Perez-Rios2010}, controllable quantum chemistry \cite{Abrahamsson2007,Quemener2016} and quantum computation \cite{Herrera2014,Karra2016}.

\begin{acknowledgments}
We thank Lewis McArd for assistance with the magnetic field setup. We acknowledge support from the UK Engineering and Physical Sciences Research Council (grant numbers EP/P01058X/1 and EP/T015241/1). KEW acknowledges  support from the Royal Society University Research Fellowship (URF\textbackslash R1\textbackslash 201134). The data presented in this paper are available from \url{http://doi.org/doi:10.15128/r1pz50gw14v}.

\end{acknowledgments}
\
\appendix*

\section{Magnetic Field Setup}
 To reach the moderately high fields at which the Feshbach resonances are predicted to occur, the magnetic bias field is generated by two pairs of coils. The first are a pair of coils which we shall refer to as the bias coils. These coils operate in the Helmholtz configuration and are used to generate the bias field necessary for the preparation of the atomic mixture. The second pair of coils are those used to generate the magnetic field gradients for the MOT and reservoir trap stages of the experiment. These coils are normally operated in an anti-Helmholtz configuration but for these measurements an H-bridge circuit is connected to the lower coil. This circuit is used to flip the direction of current flow through the coil during an experimental sequence, converting the pair to the Helmholtz configuration for the Feshbach measurement portion of the sequence. As this coil pair was initially designed to produce magnetic field gradients, operating the coil pair in the Helmholtz configuration leads to significant magnetic field curvature of 0.2~G cm$^{-2}$\ A$^{-1}$ at the position of the atoms, which we estimate to lead to a broadening of $\sim5\,\mathrm{mG}$ under typical experimental conditions. The field produced by the gradient coil pair is along the same direction as the bias coil pair and the sum of their fields is used to reach magnetic fields required to observe the Feshbach resonances. From microwave spectroscopy of the Cs transition $(f=3,m_{\rm f}=+3) \rightarrow (f=4,m_{\rm f}=+4)$, we determine the magnetic field noise to be 0.04~G rms and estimate the long-term field stability to be about 0.02~G.

\end{document}